\documentclass[notitlepage,nofootinbib,preprintnumbers,aps,superscriptaddress,twocolumn]{revtex4-1}
\usepackage{amsmath,amssymb,natbib,bm,color}
\usepackage{appendix}
\usepackage{graphicx}
\usepackage{slashed}

\definecolor{linkblue}{rgb}{0,0,0.8}
\definecolor{linkgreen}{rgb}{0,0.5,0}
\definecolor{linkdarkorange}{rgb}{1, 0.34, .2}
\definecolor{linkblack}{rgb}{0, 0, 0}
\usepackage[colorlinks,bookmarks]{hyperref}
\hypersetup{pdfpagemode=UseNone, pdfstartview=FitH, linkcolor=linkblack, citecolor=linkblack, urlcolor=linkblack}

\usepackage{verbatim}

\begin{document}
\preprint{\hbox{UTWI-01-2024, NORDITA-2024-005}}
\title{Dark Matter production during Warm Inflation via Freeze-In}
\author{Katherine Freese}
\email{ktfreese@utexas.edu}
\affiliation{Texas Center for Cosmology and Astroparticle Physics, Weinberg Institute for Theoretical Physics, Department of Physics, The University of Texas at Austin, Austin, TX 78712, USA}
\affiliation{The Oskar Klein Centre, Department of Physics, Stockholm University, AlbaNova, SE-10691 Stockholm, Sweden}
\affiliation{Nordic Institute for Theoretical Physics (NORDITA), 106 91 Stockholm, Sweden}
\author{Gabriele Montefalcone}
\email{montefalcone@utexas.edu}
\affiliation{Texas Center for Cosmology and Astroparticle Physics, Weinberg Institute for Theoretical Physics, Department of Physics, The University of Texas at Austin, Austin, TX 78712, USA}
\author{Barmak Shams Es Haghi}
\email{shams@austin.utexas.edu}
\affiliation{Texas Center for Cosmology and Astroparticle Physics, Weinberg Institute for Theoretical Physics, Department of Physics, The University of Texas at Austin, Austin, TX 78712, USA}

\begin{abstract}

We present a novel perspective on the role of inflation in the  production of Dark Matter (DM). Specifically, we explore the DM production during Warm Inflation via ultraviolet Freeze-In (WIFI). We demonstrate that in a Warm Inflation (WI) setting the persistent thermal bath, sustained by the dissipative interactions with the inflaton field, can source a sizable DM abundance via the non-renormalizable  interactions that connect the DM with the bath. Compared to the (conventional) radiation-dominated (RD) UV freeze-in scenario for the same reheat temperature (after inflation), the resulting DM yield in WIFI is always enhanced showing a strongly positive dependence on the mass dimension of the non-renormalizable operator. 
Of particular interest, for a sufficiently large mass dimension of the operator, the entirety of the DM abundance of the Universe can be  created during the inflationary phase.  For the specific models we study, we find  that the enhancement in DM yield, relative to RD UV freeze-in, is at least an order of magnitude for an operator of mass dimension 5, and as large as 18 order of magnitudes for an operator of mass dimension 10.  
Our findings also suggest a broader applicability for producing other cosmological relics, which may have a substantial impact on the evolution of the early Universe.
\end{abstract}
 
\maketitle

\textbf{\textit{Introduction.---}} An intriguing paradigm to produce dark matter (DM) is through interaction with a thermal bath where the DM abundance is established mainly by freeze-out or freeze-in. In freeze-out, the DM is in chemical equilibrium with the bath at early times. In  freeze-in, on the contrary, the DM never comes to thermal equilibrium with the bath~\cite{Hall:2009bx}. The suppressed interaction between frozen-in DM and the thermal bath can be due to renormalizable interactions with a very small coupling constant, known as infrared (IR) freeze-in~\cite{Hall:2009bx}, or because of non-renormalizable interactions with a heavy mass scale, known as ultraviolet (UV) freeze-in~\cite{Elahi:2014fsa}. While in IR freeze-in the DM abundance is set by the (low) temperature near the DM mass, in the UV freeze-in case, the abundance of DM is sensitive to the highest temperature of the bath. It is known that during reheating, the bath can achieve temperatures much higher than the reheat temperature evaluated based on the instantaneous decay of the inflaton~\cite{Chung:1998rq,Giudice:2000ex,Kolb:2003ke}. Careful consideration of the UV freeze-in of DM during reheating has shown that the DM yield can be enhanced compared to the instantaneous case~\cite{Garcia:2017tuj,Chen:2017kvz,Bernal:2019mhf,Barman:2022tzk,Garcia:2018wtq,Chowdhury:2023jft}. However, all existing studies to date on DM production via UV freeze-in have neglected the possibility that any significant amount of DM may be produced during the inflationary expansion and not be diluted away.
This however may not be the case. It is in fact possible for a persistent thermal bath  to exist throughout the inflationary epoch. This is known as the warm inflation (WI) scenario~\cite{Berera:1995ie, Berera:1995wh, Berera:1998px} and, as we will show in this \textit{letter}, it emerges as a natural framework for the efficient production of DM via freeze-in during the period of inflationary expansion. To encapsulate this novel framework we introduce the acronym WIFI, standing for `Warm Inflation  Freeze-In'. 

We find that in WIFI, the DM yield is enhanced remarkably 
compared to the conventional UV freeze-in yield from
a radiation dominated (RD) era at the same reheat temperature. As the mass dimension of the non-renormalizable operator increases, so does the enhancement; with the majority of the DM yield being generated during inflation when the mass dimension is sufficiently large. This outcome represents the pivotal contribution of our work which  provides the first complete picture of DM production via UV
freeze-in from the onset of inflation to today — unlike previous studies that treat inflation and reheating separately.

\textbf{\textit{General Setup.---\label{sec:model}}} 
In WI, the energy density of the inflaton is continuously transferred to a thermal bath through dissipative effects.  The radiation persists and eventually dominates the energy density of the Universe. This provides a smooth transition from the inflationary phase to the RD one.

In WIFI, while the inflaton
is in equilibrium with the radiation bath, it does not interact with DM. The DM, which has a mass less than the bath temperature, interacts with the radiation via a non-renormalizable interaction suppressed by some powers of a mass scale larger than the bath temperature. The suppressed interaction between DM and the bath makes DM a harmless addition to the WI framework. The evolution of the homogeneous inflaton
field $\phi$ in the presence of a thermal bath at temperature $T$ is:
\begin{align}
    &\ddot{\phi}+(3H+\Upsilon)\dot{\phi}+dV(\phi)/d\phi=0,\label{eq:inflaton} \\
    & \dot{\rho}_r+4H\rho_r=\Upsilon \dot{\phi}^2,\label{eq:rhor} \\
    & H^2=\left(\rho_\phi+\rho_r\right)/\left(3 M_{\rm{pl}}^2\right),\label{eq:Hubble}
\end{align}
where $\rho_\phi=V(\phi)+\dot{\phi}^2/2$, $\rho_r=(\pi^2/30)g_\star(T)T^4$ with $g_\star(T)$ number of relativistic degrees of freedom, and $M_{\rm{Pl}}\equiv 1/\sqrt{8\pi G}\approx 2.4 \times 10^{18}\,$GeV. 
 The dissipation term, $\Upsilon(\phi,T)$, describes the rate at which the energy in the inflaton field converts into radiation and its exact form depends on the underlying microphysical model~\cite{Berera:1998px}. 

The evolution of the number density of DM, $\chi$, denoted by $n_\chi$, is governed by the Boltzmann equation:
\begin{eqnarray}
    \dot{n}_\chi+3H n_\chi&=&T^{2n+4}/\Lambda^{2n},\label{eq:nchi}
\end{eqnarray}
where $\Lambda$ is the cutoff of the effective field theory which describes the interaction between DM and the radiation~\cite{Elahi:2014fsa}. Since, $\rho_\chi\ll \rho_r$, the contribution of DM in Eqs.~\eqref{eq:rhor} and \eqref{eq:Hubble} is safely ignored.

Recasting Eq.~\eqref{eq:nchi} in terms of the DM yield, $Y_\chi\equiv n_\chi / s$, with $s=(2\pi^2/45)g_{\star, S}(T)T^3$ being the entropy density of the thermal bath,
\begin{equation}
 Y_\chi(N_e)=\frac{45}{2\pi^2g_\star}\frac{e^{-3N_e}}{ T^3(N_e)}\int_{N_{e, 0}}^{N_e}\mathcal{I}_\chi(N'_e) dN'_e,
 \label{eq:DMyield}
\end{equation}
where
\begin{equation}
    \mathcal{I}_\chi(N_e)\equiv e^{3N_e}\frac{T^{2n+4}(N_e)}{\Lambda^{2n}H(N_e)}, \label{eq:DMYintegrand}
\end{equation}
and $N_e$ is the number of e-folds defined by $dN_e\equiv Hdt$. A vanishing DM yield at some initial time, $N_{e, 0}$, is assumed.
It is worth mentioning that $\mathcal{I}_\chi$ is the rate of change of the comoving DM number density, $N_\chi\equiv e^{3N_e}n_\chi$, i.e.  $\mathcal{I}_\chi= dN_\chi/dN_e$.

According to Eq.~\eqref{eq:DMyield}, which is applicable to UV freeze-in within any cosmology, the DM yield depends on the evolution of $T(N_e)$ and $H(N_e)$. In a WI setting, these functions can be evaluated simply from the set of Eqs.~\eqref{eq:inflaton}, \eqref{eq:rhor}, and \eqref{eq:Hubble}. Therefore, the DM yield evolution is clearly contingent on the precise details of the WI model under consideration.
Nevertheless, we can still extract overarching conclusions about the DM relic abundance within our framework. In fact, $\mathcal{I}_\chi(N_e)$ (Eq.~\eqref{eq:DMYintegrand}) is generically an exponentially increasing function through most of the inflationary phase, when $\rho_\phi\gg \rho_r$, and it becomes an exponentially decreasing function shortly after the end of inflation at the onset of the RD phase, when $\rho_\phi\ll \rho_r$. It can, therefore, be argued that the function $\mathcal{I}_\chi(N_e)$ 
is sharply peaked at some e-fold, $N_{e}^{\rm{peak}}$, which is the solution of  $d \mathcal{I}_\chi(N_e) / dN_e = 0$, i.e.,
\begin{equation}
 3+(2n+4)\frac{d\,{\rm ln}\,T(N_e)}{dN_e}-\frac{d\,{\rm ln}\,H(N_e)}{dN_e}=0.
 \label{eq:Nemax}
\end{equation}

This allows us to estimate the DM yield at late times, i.e. for $N_e> N_e^{\rm{peak}}$, by using the peak value of $\mathcal{I}_\chi$, to obtain:
\begin{eqnarray}
   \nonumber Y_\chi(N_e)&\simeq&\frac{45}{2\pi^2g_{\star}}\frac{e^{3\left(N_{e}^{\rm peak}-N_e\right)}}{\Lambda^{2n}T^3(N_e)}\Delta N_{e}^{\rm peak} \\
   &\times& \frac{T^{2n+4}(N_e^{\rm peak})}{ H(N_e^{\rm peak})}, \quad (N_e> N_e^{\rm{peak}}),
   \label{eq:estY}
\end{eqnarray}
where $\Delta N_e^{\rm{peak}}\gtrsim 1$ denotes the
half-width of $\mathcal{I}_\chi$.
To better understand the importance of this result, one can compare it with the final yield of DM produced from UV freeze-in (Eq.~\eqref{eq:nchi}) in a RD epoch which starts at some initial time $N_{e, 0}$ with temperature $T(N_{e, 0})=T_{\rm rh}$:
\begin{equation}
Y^{\rm{RD}}_{\chi,\infty}(T_{\rm rh})\simeq \frac{1}{\sqrt{2}}\left(\frac{45}{\pi^2 g_\star}\right)^{3/2}\frac{1}{2n-1}\frac{M_{\rm{Pl}}T_{\rm{rh}}^{2n-1}}{\Lambda^{2n}}. 
    \label{eq:YDMRD}
\end{equation}
While the yield of the DM in UV freeze-in from a RD era depends on the highest temperature, i.e. $T_{\rm{rh}}$, and freezes in very quickly to its final value, in the WI scenario, the bulk of it is produced at $N_e^{\rm peak}$, given by Eq.~\eqref{eq:Nemax}. 

Since in WIFI, DM production starts from the beginning of the inflation and always peaks before the onset of the RD phase, the final yield of DM is expected to be larger than the corresponding value obtained in UV freeze-in from the RD scenario. 
To make a proper comparison between these two cases, for a given WI model and its corresponding temperature evolution, we define the reheat temperature, $T_{\rm rh}$, as the bath temperature once the Universe enters the RD phase, i.e. $T_{\rm{rh}}\equiv T(\epsilon_H=2)$ where $\epsilon_H\equiv-\Dot{H}/H^2$. Then, we introduce:
\begin{equation}
    R^{(n)}_{\chi}\equiv Y_{\chi,\infty}/Y^{\rm{RD}}_{\chi,\infty}(T_{\rm{rh}}),\label{eq:enhancement_ratio}
\end{equation}
as the ratio of $Y_{\chi,\infty}$, the final value of the yield obtained from Eqs.~\eqref{eq:DMyield} and \eqref{eq:DMYintegrand}, to $Y_{\chi,\infty}^{\rm{RD}}(T_{\rm{rh}})$, evaluated from Eq.~\eqref{eq:YDMRD}.

By taking the limit of the semi-analytical expression for the yield given by Eq.~\eqref{eq:estY} when $N_e\rightarrow\infty$, we can estimate $R^{(n)}_{\chi}$ as:
\begin{eqnarray}
    R_\chi^{(n)}\simeq (2n-1) \frac{\mathcal{I}_\chi (N_e^{\rm{peak}})}{\mathcal{I}_\chi (N_e^{\rm{RD}})}\Delta N_e^{\rm{peak}}, \label{eq:boost_SA}
\end{eqnarray}
where $N_e^{\rm{RD}}$ corresponds to the e-fold at which the Universe becomes RD, i.e. $N_e^{\rm{RD}}\equiv N_e(\epsilon_H=2)$.
Evidently, from Eq.~\eqref{eq:boost_SA} it follows generically that $R_\chi^{(n)}\gg 1$, i.e. production of DM in WIFI is always enhanced compared to the one in a RD phase. This enhancement increases noticeably with $n$ mainly because the larger the value of $n$, the earlier in time $\mathcal{I}_\chi$ reaches its peak. This causes the onset of the RD phase ($N_e^{\rm{RD}}$) to move further away from the peak ($N_e^{\rm{peak}}$), which in turn leads to a significant enhancement of the ratio $\mathcal{I}_\chi (N_e^{\rm{peak}})/\mathcal{I}_\chi (N_e^{\rm{RD}})$.

 \textbf{{\textit{A Representative Example.---}}} To illustrate the efficiency and richness of this novel mechanism, let us consider the case of a quartic inflaton potential $V(\phi)=\lambda \phi^4$ and a linear dissipation rate $\Upsilon\propto T$. This is motivated by 
 Warm Little Inflaton~\cite{Bastero-Gil:2016qru, Bastero-Gil:2019gao}, which constitutes the simplest successful realization of WI, in agreement with the cosmic microwave background (CMB) data.

 Central to our analysis is the the effectiveness at which the inflaton converts into radiation, parameterized by the dimensionless parameter: 
\begin{equation}
    Q\equiv \Upsilon/(3H).
\end{equation}
 We specifically consider two
 scenarios with initial dissipation strength  equal to $Q_0=10^{-2}$ and $Q_0=1$. In both cases, we assume 60 e-folds of inflation, the Standard Model (SM) value for $g_\star=106.75$ and a self-interacting quartic coupling $\lambda$ equal to $1.15\times10^{-15}$ and $7.91\times10^{-16}$, respectively, in order to reproduce cosmological observables in agreement with the CMB bounds set by \textit{Planck}~\cite{Planck:2018jri}. Specifically for $Q_0=10^{-2}$ ($Q_0=1$), we get  $n_s\simeq0.967$ ($n_s\simeq 0.968$) and $r\simeq7.6\times 10^{-3}$ ($r\simeq3.9\times 10^{-4}$)~\cite{Montefalcone:2023pvh}.  
 
\begin{figure}[ht!]
    \centering
    \includegraphics[width=\linewidth]{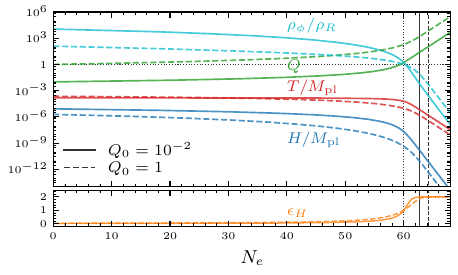}
    \caption{
    The evolution of various quantities for the case of WI with $V(\phi)=\lambda\phi^4$, for two initial values of the dissipation strength, $Q_0=10^{-2}$ (solid) and $Q_0=1$ (dashed). The dotted vertical line shows the end of inflation.
    The onset of radiation domination is marked by a vertical line (respectively solid and dashed for $Q_0=10^{-2}$ and $Q_0=1$).
    }
    \label{fig:dynamics}
\end{figure}

 In Fig.~\ref{fig:dynamics}, we illustrate the cosmological dynamics  for these two examples. The inflaton potential energy
dominates the energy balance throughout the inflationary phase, 
while sustaining a radiation bath with nearly constant temperature 
above the corresponding Hubble rate. In both cases, the dissipation becomes strong ($Q>1$) towards the end of inflation, allowing for the radiation to smoothly take over as the dominant component  within roughly 2 (4) e-folds after the end of inflation for $Q_0=10^{-2}$ ($Q_0=1$). Following this transition, both the temperature and Hubble rate settle into their standard RD Universe evolution, i.e. $T\sim e^{-N_e}$ and $H\sim T^2$.

\begin{figure*}[ht!]
    \centering
    \includegraphics[width=\linewidth]{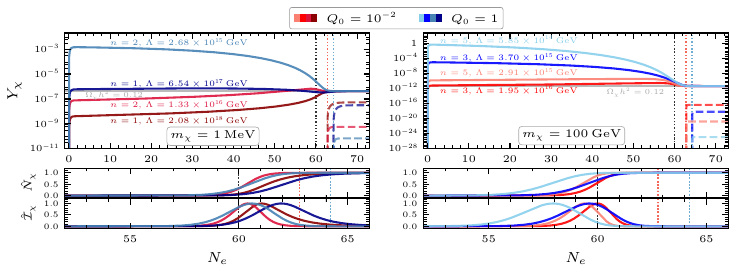}
    \caption{\textit{Top:} 
    The DM yield $Y_\chi$ as a function of the number of e-folds (solid lines), assuming vanishing initial DM abundance, and given the background dynamics shown in Fig.~\ref{fig:dynamics} for $Q_0=10^{-2}$ (red) and $Q_0=1$ (blue). In each case, we show the resulting yield for various values of $n$, with the DM mass $m_\chi$ and the scale $\Lambda$ fixed to match the observed DM relic abundance $\Omega_{\rm{CDM}}h^2=0.12$. In both panels, the dashed lines represent the corresponding yield evolution in the conventional RD UV freeze-in. The vertical lines are the same as in Fig.~\ref{fig:dynamics}.
    \textit{Bottom:}  The comoving DM number density $N_\chi\equiv e^{3N_e}n_\chi$ and its derivative $\mathcal{I}_\chi=dN_\chi/dN_e$, corresponding to the yield evolution in the top panel and scaled so that their maximum equals 1.}
    \label{fig:yield_DM}
\end{figure*}

Our main results are illustrated in Fig.~\ref{fig:yield_DM}. On the top panel of the figure, we present the full evolution of the DM yield for $Q_0=10^{-2}$ and $Q_0=1$, depicted in varying shades of red and blue respectively, each corresponding to different values of $n$. For all individual cases analyzed, solid lines represent the evolution assuming a vanishing abundance at the start of inflation, and matched to the observed DM relic abundance of $\Omega_{\rm{CDM}}h^2=0.12$~\cite{Planck:2018vyg}. Dashed lines instead depict the corresponding DM yield evolution under the assumption of a vanishing abundance at the onset of the RD epoch. Specifically, on the left (right) panel, we take $m_\chi=1\,$MeV ($m_\chi=100\,$GeV) and show the cases $n=\{1,2\}$ ($n=\{3,5\}$). 

Notably the DM relic abundance obtained from the full background evolution
is always significantly greater than the corresponding yield from the subsequent RD era. Specifically, the enhancement is at least 1 order of magnitude and it increases exponentially with $n$, corroborating  the general conclusions established in the previous section,  Eq.~\eqref{eq:boost_SA}.  In addition, we also note that the enhancement in the DM yield is consistently higher for $Q_0=1$ compared to $Q_0=10^{-2}$. This can be understood from the the sharper temperature drop near the end of inflation and longer transition to RD phase of the $Q_0=1$ case, as evidenced in Fig.\ref{fig:dynamics}. These factors cause an earlier peak in $\mathcal{I}_\chi$ and an increased distance between this peak and the onset of RD phase, thereby leading to a higher enhancement ratio $R^{(n)}_\chi$ for $Q_0=1$ relative to $Q_0=10^{-2}$.

The insights gained from the top panel of Fig.~\ref{fig:yield_DM} are further elucidated in its lower panels, which illustrate the corresponding evolution of the comoving DM number density $N_\chi$ and its derivative $\mathcal{I}_\chi$ (Eq.\eqref{eq:DMYintegrand}), both scaled so that their maximum is $1$. From the figure, we observe that the critical increase in $N_\chi$ always occurs at the peak of $\mathcal{I}_\chi$,  and within a few e-folds around it. In all cases, the relic value of $N_\chi$ is predominantly established before the RD epoch,  which demonstrates why our mechanism consistently leads to an enhancement in DM abundance compared to the RD UV freeze-in. Furthermore, the panel showing the $\mathcal{I}_\chi$ evolution visually confirms our expectation that a greater enhancement in DM yield will correspond to a larger gap between the peak of $\mathcal{I}_\chi$ and the onset of the RD phase. This explains the more pronounced boost in DM production with increasing values of $n$, as well as the amplification for the $Q_0=1$ case over $Q_0=10^{-2}$. Of particular interest is the behavior for the largest  $n$ values analyzed here, specifically within the $Q_0=1$ scenario. In these instances, we observe that essentially all of the DM is produced during the inflationary phase itself, leading to the greatest enhancement in DM yield relative to the RD UV freeze-in.  The possibility of producing all of the DM in the Universe during the inflationary period itself is a novel consequence of the WIFI scenario.

\begin{figure}[hb!]
    \centering
    \includegraphics[width=1\linewidth]{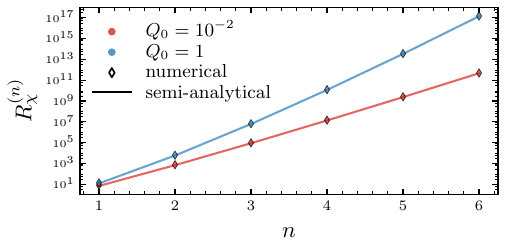}
    \caption{The
    ratio $R_\chi^{(n)}$ (Eq.~\eqref{eq:enhancement_ratio}) as a function of $n$
    for the two WI scenarios with $Q_0= 10^{-2}$ (red) and $Q_0 = 1$ (blue), both the numerical results (diamonds) and the corresponding semi-analytical estimates (solid lines).}
    \label{fig:boost}
\end{figure}

Fig.~\ref{fig:boost} illustrates the potentially enormous efficacy of DM production via WIFI.
In this figure, the enhancement ratio $R^{(n)}_\chi$, Eq.~\eqref{eq:enhancement_ratio}, is plotted as a function of $n$ for both scenarios: $Q_0=10^{-2}$ (red) and $Q_0=1$ (blue).  The numerical evaluations of $R^{(n)}_\chi$ are shown as diamonds, while our semi-analytical estimates, Eq.~\eqref{eq:boost_SA}, are represented by solid lines. As anticipated, $R^{(n)}_\chi$ is always greater than $1$ and it increases exponentially with $n$. Specifically, $R^{(n)}_\chi\gtrsim 10$ even for $n=1$ and $R^{(n)}_\chi\gg 10^{3}$ for all $n\geq 3$. This underscores the remarkable efficiency of DM production from WIFI, far surpassing the yield from the conventional RD epoch subsequent to WI. In addition, the consistently larger values of $R^{(n)}_\chi$ for the $Q_0=1$ scenario compared to $Q_0=10^{-2}$ emphasize the substantial impact of the specific WI dynamics on the overall enhancement in DM yield within our framework. While higher $Q$ results in enhanced DM production in our study, this trend does not hold across all WI constructions.

In all of these cases, our semi-analytical estimates closely match the numerical evaluations of $R^{(n)}_\chi$, with discrepancies consistently below 15\%. 

It is important to comment on the allowed DM mass range. Beyond the simple upper bound on $m_\chi$ set by the
reheating temperature,
matching the DM relic abundance introduces a notable degeneracy between the scale 
$\Lambda$, the exponent
$n$ of the non-renormalizable operator, and $m_\chi$. Further considerations within our framework, namely the requirement that DM is cold ($m_\chi\gtrsim\,{\rm keV}$) and never reaches thermal equilibrium with the bath, 
narrow down the viable parameter space. As a result, the minimum DM mass required to match the observed abundance today tends to increase with $n$, reflecting the nuanced interplay between $\Lambda$, $n$ and $m_\chi$ within our framework.\footnote{For the interested reader, we refer to Fig.~\ref{fig:4} in the Supplementary Material which displays the viable parameter space in the $\Lambda$-$m_\chi$ plane for the two WI scenarios presented in this work.}

Lastly, despite the inflationary origin of the WIFI framework, it does not produce DM isocurvature perturbations. In freeze-in, DM energy transfer and pressure depend solely on the thermal bath temperature, causing DM to inherit the adiabatic perturbations generated via WI~\cite{Strumia:2022qvj,Racco:2022svs,Bellomo:2022qbx,Taylor:2000ze}.

\textit{\textbf{Discussion and Summary.---}\label{sec:Dis}} In this \textit{letter}, we have introduced a novel perspective on the role of inflation in the thermal production of DM via freeze-in. We demonstrated that in a WI setting the persistent thermal bath can source a sizable DM abundance via the non-renormalizable interaction characteristic of UV freeze-in, Eq~\eqref{eq:nchi}. Compared to the (conventional) RD UV freeze-in scenario for the same reheat temperature, the resulting DM yield in WIFI is always enhanced,  showing a strongly positive dependence on the exponent $n$ of the non-renormalizable operator, see Eq.~\eqref{eq:boost_SA}. 
Through a physically-motivated example, we illustrated the efficacy of DM freeze-in from WI, as well as the significant influence of the specific
WI dynamics on the overall enhancement in DM yield, see Figs.~\ref{fig:yield_DM} and \ref{fig:boost}. Notably, even for a dimension-five operator, we get an increase in the DM yield of at least
an order of magnitude a novel contribution that has not been achieved by other non-standard cosmologies~\cite{Bernal:2019mhf}. Additionally, for sufficiently large values of $n$, the DM relic abundance is entirely created during inflation, leading to the greatest enhancement in DM yield relative to the RD UV freeze-in.  This new central role of the inflationary phase in the DM production highlights a key distinction of our mechanism from RD UV freeze-in: in WIFI, the relic DM yield is mostly determined by the specific period where the interplay of bath temperature and Hubble parameter fulfills Eq.~\eqref{eq:Nemax}, rather than by the highest temperature of the bath alone. 

Furthermore, the WIFI mechanism proposed in this \textit{letter} emerges as a natural extension of the WI framework, which often generates a bath of beyond the SM particles~\cite{ Berera:2008ar,Berera:2002sp, Moss:2006gt, Bastero-Gil:2016qru, Berghaus:2019whh}. These scenarios necessitate new particles and interactions to eventually produce the SM. Our framework leverages this by considering one of these new fields as DM, interacting with the radiation through heavy mediators. This  highlights our findings' relevance within a WI setting and also suggests a broader applicability, such as producing other  cosmological relics, potentially crucial in the early Universe evolution~\cite{Taylor:2000jw}. 
We leave the study and exploration of concrete realizations to future work.

\textit{\textbf{Acknowledgments.---}\label{sec:Acknowledgments}} K.F.\ is Jeff \& Gail Kodosky Endowed Chair in Physics at the University of Texas at Austin, and K.F.\ and G.M.\  are grateful for support via this Chair. K.F.\ G.M.\ and B.S.E.\  acknowledge support by the U.S.\ Department of Energy, Office of Science, Office of High Energy Physics program under Award Number DE-SC-0022021. K.F.\ and G.M.\ also acknowledge support from the Swedish Research Council (Contract No.~638-2013-8993). The authors would like to thank the Leinweber Center for Theoretical Physics, University of Michigan, for hospitality while this work was being completed. We acknowledge the use of \texttt{WarmSPy}~\cite{Montefalcone:2023pvh}, and the Python packages \texttt{Matplotlib}~\cite{matplotlib}, \texttt{Numpy}~\cite{numpy} and \texttt{Scipy}~\cite{scipy}.
\bibliographystyle{apsrev4-1}
\bibliography{letter}

	\clearpage
	
	\onecolumngrid
\begin{center}
  \textbf{\large Supplementary Material for Dark Matter production during Warm Inflation via Freeze-In}\\[.2cm]
  \vspace{0.05in}
  {Katherine Freese, Gabriele Montefalcone, and Barmak Shams Es Haghi}
\end{center}

\begin{figure*}[ht!]
    \centering
    \includegraphics[width=1\linewidth]{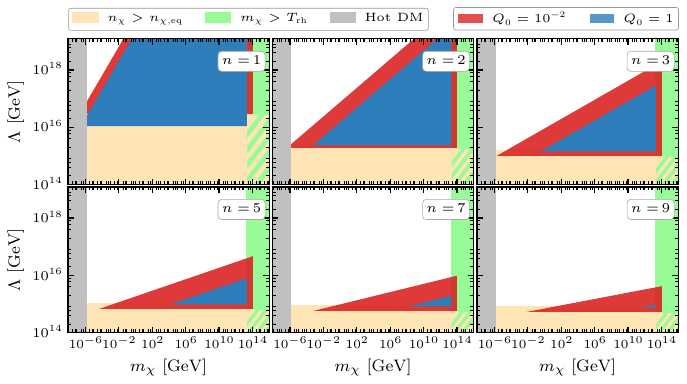}
    \caption{Constraints on DM production from WIFI in the $(m_\chi,\Lambda)$ plane, with $m_\chi$ denoting the DM mass and $\Lambda$ representing the scale of the non-renormalizable operator connecting DM with the bath, Eq.~\eqref{eq:nchi}. Displayed in separate panels for $n=\{1,2,3,5,7,9\}$, where $n$ indicates the exponent in the non-renormalizable interaction. The gray, yellow, and green shaded areas are excluded by structure formation, the requirement that DM does not come in to thermal equilibrium with the bath, and the temperature of the radiation respectively. Within the red and the blue shaded regions, DM is overproduced $(\Omega_\chi> \Omega_{\rm CDM})$
    for the WI scenarios with $Q_0=10^{-2}$ and $Q_0=1$, respectively, each featuring a quartic inflaton potential and a temperature-linear dissipation rate. Here, $\Omega_{\rm CDM}$ stands for the observed (cold) DM abundance.}
    \label{fig:4}
\end{figure*}
\end{document}